%% file: dist_mem_ext_sort.tex
\documentclass[10pt,conference,letterpaper]{IEEEtran}
\usepackage{times,amsmath,epsfig}
\usepackage{color}

\newtheorem{theorem}{Theorem}[section]

\newcommand{\sq}{\hbox{\rlap{$\sqcap$}$\sqcup$}}
\newcommand{\qed}{\hspace*{\fill}\sq}

\usepackage{amssymb}
\usepackage{url}

\newcommand{\CanonicalMergesort}{\textsc{CanonicalMergeSort}}

\input{makros.tex}
\renewcommand{\frage}[1]{}

\title{Scalable Distributed-Memory External Sorting}

\author{%
{Mirko Rahn, Peter Sanders, Johannes Singler\small$^4$}%
\vspace{1.6mm}\\
\fontsize{10}{10}\selectfont\itshape
Karlsruhe Institute of Technology\\
\fontsize{10}{10}\selectfont\itshape
Postfach 6980, 76128 Karlsruhe, Germany\\
\fontsize{9}{9}\selectfont\ttfamily\upshape
rahn@ira.uka.de sanders@kit.edu singler@kit.edu%
}
\thanks{Partially supported by DFG grant SA 933/3-2.}

\begin{document}

\maketitle

\begin{abstract}
  We engineer algorithms for sorting huge data sets on massively parallel
  machines. The algorithms are based on the multiway merging paradigm.  We first outline an algorithm whose I/O requirement is close to a lower bound. Thus,
  in contrast to naive implementations of multiway merging and all other
  approaches known to us, the algorithm works with just two passes over the data
  even for the largest conceivable inputs.  A second algorithm reduces
  communication overhead and uses more conventional specifications of the result
  at the cost of slightly increased I/O requirements. An implementation wins the 
  well known sorting benchmark in several categories and by a large margin over its
  competitors.
\end{abstract}

\bigskip

\footnotetext[4]{Partially supported by DFG grant SA 933/3-2.}

\section{Introduction}
There are currently two main ways to handle huge inputs in a cost-efficient
manner: keeping most data externally on low cost hard disks, and clustering many
inexpensive machines.  The combination of both approaches allows relatively
cheap machines to handle huge inputs that would otherwise require high-end,
power hungry super-computers with lots of internal memory.
On high-end machines equipped with sufficient disk
bandwidth, one could handle inputs of unprecedented size. For example, a mid-size cluster
with 1024 Terabyte disks which cost about 100~KEuro, can scan a Petabyte of
data in a few hours.

Perhaps the most important nontrivial operation needed for processing such huge
data sets is sorting. For example, sorting (or similar computations) can be used
to build index data structures or to arrange geometrical data such that close-by
data can be processed together (e.\,g., using space filling curves).
Fundamental lower bounds \cite{AggVit88} basically tell us that in order to
process inputs significantly larger than the \emph{cumulative} main memory size
$M$~\footnote{To avoid cumbersome notation, we will also use $M$ to denote the size of a run in external mergesort algorithms. Depending on details of the implementation, the run size might be a factor around two smaller or larger \protect\cite{Knu98}. However, this difference has little effect on the overall performance.}, at least two passes\footnote{One pass comprises reading and writing the data once.} over the data are needed. More precisely, up to $M^2/B$
elements can be processed in two passes (refer to Table~\ref{t:symbols} for explanations of the symbols $P,M,D,B,N,R$ used in this paper).

Although there is a lot of previous work on parallel external sorting,
the problem is not solved yet. In particular, algorithms used in practice
can have very bad behavior for worst-case inputs, whereas all previous theoretical results lead to algorithms that need more than two passes even for easy inputs.
Section~\ref{s:related} gives more details.

In Section~\ref{s:theoretical}, we outline a
conceptually simple variant of multiway mergesort that needs two passes even for inputs whose size is close to the theoretical limit for being sorted with two passes. However, this algorithm has relatively large communication overhead and
outputs the data in a globally striped fashion, i.\,e., subsequent blocks of
output are allocated on subsequent PEs (processing elements). Therefore, in
Section~\ref{s:practical}, we refine the algorithm so that it needs very little
communication and outputs the data in a format more conventional in parallel
computing, and more convenient for further processing: PE $i$ gets
the elements of ranks $(i-1)N/P+1,\ldots,iN/P$ where $N$ is the total number of
elements and $P$ is the number of PEs\footnote{Depending on the implementation, $P$ might denote the number of nodes, each of which could have several cores.}. At least on the average, and up to small ``clean up'' costs,
this \CanonicalMergesort\ algorithm needs only two passes and communicates elements only once.
Section~\ref{s:experiments} gives experimental results on a careful implementation described in Section~\ref{s:implementation}. These experiments show that the algorithm performs very well in practice
An in-place implementation sorts about 564\,GB/min with 195 8-core nodes and 780~disks,
leading the ``Indy GraySort'' category of the SortBenchmark~\cite{SortBenchmarkHomepage} in 2009.
We summarize the results and outline possible future work in Section~\ref{s:conclusion}.

\begin{table}[b]
\caption{\small \sf \textbf{Symbols used in this paper. We generally omit trivial rounding issues when dividing these quantities.}\label{t:symbols}}
\begin{center}
\begin{tabular}{|l|l|}
\hline
Resource/Number & Symbol \\
\hline
\hline
\#PEs & $P$ \\
\hline
internal memory (in \#elements) & $M$ \\
\hline
\#disks & $D$ \\
\hline
block size (in \#elements) in the EM model & $B$\\
\hline
\#elements & $N$ \\
\hline
\#runs & $R$ \\
\hline
\end{tabular}
\end{center}

\end{table} 

\section{Related Work}\label{s:related}

Since sorting is an essential ingredient of most external memory 
algorithms, considerable work has been invested in finding I/O-optimal parallel
disk sorting algorithms (e.\,g.,
\cite{VitShr94both,NodVit95,BarGroVit97,HutSanVit05}) that
approach the lower bound of $2N/DB(1+\lceil\log_{M/B}N/M\rceil)$ I/O operations
for sorting $N$ elements on a machine with $D$ disks, fast memory size\footnote{All sizes are given in number of elements.}
$M$ and block size $B$. The challenge is to avoid getting a base
$M/DB$ for the logarithm that spoils performance for very large
systems.  An important motivation for this paper is the observation that a
large $D$ only makes sense in a system with many processors.  Although
\cite{VitShr94both,NodVit93,AggPla94} develop sophisticated 
asymptotically optimal parallel algorithms,
these algorithms imply considerable constant factors of overhead with respect
to both I/Os and communication compared to the best randomized sequential
algorithms \cite{BarGroVit97,HutSanVit05,DemSan03}.

We only note in passing that there is also considerable work on parallel disk
sorting with shared-memory parallel processors, e.\,g., \cite{NBCGL94}. This is an
easier problem since communication overhead is less of an issue.

Many external memory algorithms for distributed-memory machines have been
proposed (e.\,g., see the references in \cite{Ric94}).
One of the most successful ones is NOW-Sort \cite{AACHP97} which is
somewhat similar to our algorithm \CanonicalMergesort, i.\,e., it sorts up to $M^2/(PB)$
elements in two passes. However, it only works efficiently for random inputs. In
the worst case, it deteriorates to a sequential algorithm since all the data
ends up in a single processor. This problem can be repaired by finding
appropriate splitter keys in a preprocessing step \cite{MRL98}. However, this
costs an additional scan of the data and still does not result in exact
partitioning.

In \cite{Rajasekaran04}, a merge-based parallel external sorting algorithm is
proposed that is inspired by parallel mesh algorithms. This algorithm needs at
least four passes over the data.%
\footnote{This bound is derived from the bounds in the paper assuming that 
 logarithms with fractional values have to be rounded up.}

In \cite{ChaCor06} an algorithm based on column-sort is proposed that sorts up
to $(M/P)^{3/2}/\sqrt{2}$ elements using three passes over the data. Using one additional pass,
the input size can be increased to $\max(\Oh{M^{3/2}, (M/P)^{5/3}})$ elements.  It
is instructive to use some realistic numbers. On current machines it is quite
realistic to assume about 2\,GiB of RAM per core. Using this number, the amount
of data that can be sorted with the three pass algorithm is limited to inputs of
size around ${2^{31}}^{3/2}/\sqrt{2}=2^{46}$, i.\,e. about 64\,TiB regardless
of the number of available PEs.

In \cite{DDHM02}, a general emulation technique for emulating parallel algorithms
on a parallel external memory machine is developed. It is proposed to apply this
technique to a variant of sample sort.  This results in an algorithm that needs
five passes over the data for sorting $\Oh{M^2/(PB)}$ elements.

Our exact partitioning algorithm for parallel multiway merging owes a lot to
\cite{VSIR91} where multiway merging is used for shared-memory parallel sorting.

\section{Mergesort with Global Striping\label{s:theoretical}}

Since multiway mergesort is a good algorithm for parallel disk external sorting
and parallel internal sorting, it is a natural idea to use it also for parallel
external sorting. Here we outline how to do this in a scalable way:
The first phase is \emph{run formation} where initial runs of size $M$ are loaded into the \emph{cumulative} memory of the parallel machine,
sorted \emph{in parallel}, and written back to disk. 

Next follow one or more merging phases\footnote{In general we need   $\lceil\log_{\Th{M/B}}\frac{N}{M}\rceil$ merging phases.} where up to $k=\Oh{M/B}$ sorted runs are merged in a single pass.  The challenge is that we are only allowed a constant number of buffer blocks for each run. In particular, we may not be able to afford $k$ buffer blocks on \emph{every} PE. We solve this by fetching a batch of $\Th{M/B}$ blocks at a time into the internal memory (those blocks that will be needed next in the merging process), extracting the $\Th{M}$ smallest unmerged elements using internal parallel merging, and writing them to the disks.  Fetched elements that are larger than the smallest unfetched elements are kept in internal memory until the next batch. Note that this is possible since by definition of the blocks to be fetched, for each run, at most $B$ elements remain unmerged.  Note that we could even afford to replace batch merging by fully-fledged parallel sorting of batches without performing more work than during run formation.

The difficult part is how to do the disk accesses efficiently.  However, this
can be done in an analogous fashion to previous (sequential) parallel disk
sorting algorithms \cite{BarGroVit97,DemSan03,HutSanVit05}.  The runs and the
final output are \emph{striped} over all disks, i.\,e., subsequent blocks are
allocated on subsequent disks.  This way, writing becomes easy: We maintain $D$
buffer blocks.  Whenever they are full, we output them to the disks in parallel.
Efficiently fetching the data is more complicated.  A \emph{prediction sequence}
consisting of the smallest element in each data block can be used to predict in
which order the data blocks are needed during merging \cite{Knu98,DemSan03}.
Using randomization, some buffer space, and appropriate prefetching
algorithms, it is then possible to make good use of all disks at once. The only
part of this algorithm that is not straight-forward to parallelize is the
prefetching algorithm. In
Appendix~\ref{a:theodetails}, we outline an efficient 
prefetching algorithm for the cases $B=\Om{\log P}$ and $M=\Om{DB\log D}$.

We believe that the above algorithm could be implemented efficiently. However, it
requires a substantial amount of communication: During run formation, all the
data has to be communicated in the parallel sorting routine and again for
writing it in a striped fashion. Similarly, during a merging pass, the data has
to be communicated during internal memory multiway merging and for outputting
it in a striped fashion. Moreover, globally striped output is often not what is needed
for further processing so that we need 4--5 communications for two passes of
sorting. In the next section we want to bring this down to a single
communication at least in the best case.

\section{\CanonicalMergesort\label{s:practical}}

In the following, we describe a variant
of parallel external mergesort that produces its output in way more canonical for parallel processing -- PE $i$ gets
the elements of ranks $(i-1)N/P+1,\ldots,iN/P$ and this data is striped over the local disk. This is not only more useful for some applications but it 
also reduces the amount of communication to a minimum at the price of some additional I/Os.

In the first phase, $R=N/M$ global runs of size $M$ (the last run might be smaller) are created.
Section~\ref{ss:formruns} gives more details.
This is similar to the algorithm of Section~\ref{s:theoretical} but now the output is not striped globally over the disks but output locally, which saves communication.  Moreover, if all runs have a similar input distribution, most elements will already end up on the PE where they are needed for a globally sorted final output.  In order to make this assumption approximately true, each PE chooses its participating blocks for the run randomly.  This is implemented by randomly shuffling the IDs of the local input blocks in a preprocessing step.

In the second phase, multiway selection operations are performed on all runs.
In general, a multiway selection operation finds the element $e$ with global rank $r$ from $R$ sorted sequences, and returns $R$ splitter positions which partition the sequences with respect to $e$.
For external parallel mergesort, each PE $i$ selects for each run the first element it is supposed to contain in the final result, resulting in $P-1$ splitter elements per run.
After communicating the splitter positions to PEs $i$ and $i-1$, every PE knows the elements it has to merge (see Section~\ref{ss:select} for more details).

The data is then redistributed accordingly using a global external all-to-all operation described in Section~\ref{ss:allall}.
If the input is uniformly distributed, or if global randomization is applied, most of the data will already be in the right place, so the all-to-all operation takes only little time.

In the third phase, the data is merged locally.
Each element is read and written once, no communication is involved in this phase.
The internal computation  amounts to $\Oh{N/P \log R} = \Oh{N/P \log N/M}$.
Overall, we need $\Oh{N/P\, \log N}$ internal computation, with a very low constant factor.

\begin{figure*}
 \includegraphics[width=\textwidth]{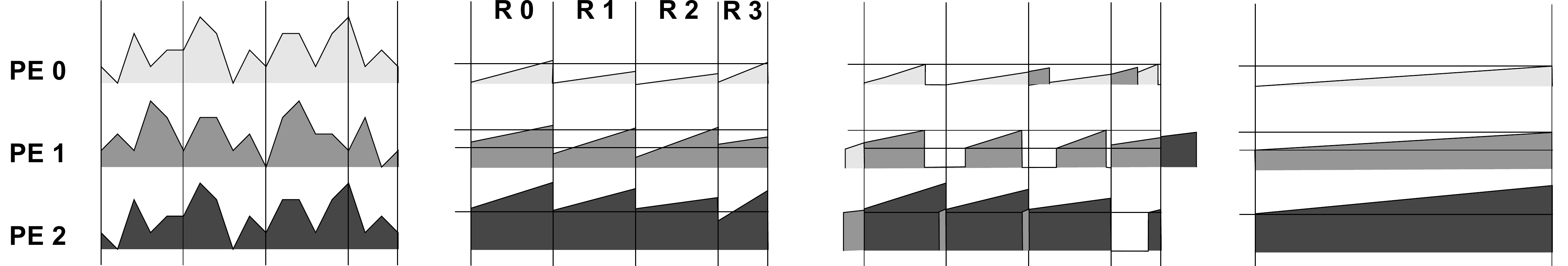}
\caption{\small \sf \textbf{\CanonicalMergesort: On the very left the initial input situation, going to the right the results of the three phases: run formation, redistribution (hopefully negligible), local merging. The y-axis denotes the element rank, the horizontal lines illustrate the global splitters.}\label{f:canonicalmergesort}}
\end{figure*}

An overview of \CanonicalMergesort\ can be found in Figure~\ref{f:canonicalmergesort}.

\subsection{Multiway Selection}\label{ss:select}

As stated before, a multiway selection finds out the element $e$ with global rank $r$ among $R$ sorted sequences, plus the splitter positions that partition the sequences with respect to $e$.
Let the length of the sequences $M$ be a power of two, rounded up and (conceptually) fill up with $\infty$ otherwise.

We maintain approximate splitter positions that are moved in steps of size $s$.  The basic algorithm uses initial splitter positions $0$ and step size $s=M$.  Within a \emph{round}, the splitter corresponding to the smallest element is increased by $s$ until the number of elements to the left of the splitters becomes larger than $r$. Then, $s$ is halved and the splitters corresponding to the largest element are decreased by $s$ while the number of elements to the left of the splitters is still larger than $r$.  This process is repeated until $s=1$.  After at most $\ceil{\log_2 M}$ rounds, the process terminates. Since in each half round every splitter is touched at most once, the overall number of sequence elements touched is $\Oh{R\log M}$ and the total execution time is $\Oh{R\log R\, \log M}$ using a priority queue for identifying the sequences to be touched.

In phase two of our algorithm, processor $i$ runs multiway selection for rank $r=iN/P$. Although these selections can run in parallel, they have to request data from remote disks and thus the worst case number of I/O steps is $\Oh{RP\log M}$ when a constant fraction of request is directed to a single disk. This could be a bottleneck for large $P$.  
This problem is greatly reduced by the randomization used during run formation. 
Furthermore, during run formation, we store every $K$-th element of the sorted run as a sample (for some parameter $K$). During multiway selection, this sample is used to find initial values for the approximate splitters. As a third optimization, we cache the most recently accessed disk blocks to eliminate the $R\log B$ last disk accesses of a multiway selection. In our implementation, we keep the sample in main memory. In our experiments, the resulting selection algorithm takes negligible time. In Appendix~\ref{app:select} we analyze a slightly more complicated variant that provably scales to very large machines, still using only very little time.

\subsection{Internal Memory Parallel Sorting}\label{ss:formruns}

We essentially use a distributed-memory implementation of the parallel multiway merging approach already used in \cite{VSIR91,SSP07}. We therefore only outline
the algorithm.
Each node sorts its local data.
Then, the internal memory variant of the multiway selection algorithm from Section~\ref{ss:select} is used to split the $P$ sorted sequences 
into $P$ pieces of equal size. An all-to-all communication is used to
move the pieces to the right PE. Note that in the best case, this is the only time when the data is actually communicated.

\subsection{External All-to-All}\label{ss:allall}

In an \emph{all-to-all} operation, each PE sends and receives different amounts of data to/from all other PEs.
Compared to the ordinary all-to-all operation provided by MPI we are facing two problems.
First, each PE might have to communicate more data than fits into its local memory. We solve this problem by splitting the external all-to-all into $k$  internal memory suboperations by logically splitting the data sent to a receiver into $k$ (almost) equally-sized parts.
The choice of $k$ depends on the available internal memory but will be at most  $\Oh{R}$. The second problem is that the data has to be collected from $R$ different runs. We therefore assemble the submessages by consuming all the 
participating data of run $i$ before switching to run $i+1$. This way, 
each PE $j$ needs only a single buffer block for each PE that it sends data to.  Note that due to randomization, the number $P'$ of required blocks will grow much more slowly than the worst case of $P-1$ communication partners.
The total number of I/O steps for data volume $V$ will be $\frac{2V}{PB}+\Oh{RP'}$.

\subsection{Summary of the Analysis\label{ss:analysis}}

The most easy summary of the analysis is that \CanonicalMergesort\ needs
I/O volume $4N + \oh{N}$, communication volume $N + \oh{N}$, and local work similar to
a fast sequential internal algorithm. Here, the ``$\oh{\cdot}$''-notation expresses
that the overheads are independent of the input size $N$ or only grow sublinearly. A little more care must be taken however, since these bounds only hold under a number of assumptions on the values of the other machine parameters, namely $P$, $M$, $B$, and $D$. To simplify matters a little bit, we assume that $D=\Th{P}$. We also introduce the shorthand
$m$ for the local memory size $M/P$. For example, on our machine, we have
$P\in 1..200$ nodes (with 8 cores each), $D=4P$, $m=2^{34}$ byte, and $B=2^{23}$ byte.

The theoretically 
most important restriction is that
the maximal amount of data that can be sorted is $\Oh{M\cdot\frac{M}{PB}}=\Oh{P\frac{m^2}{B}}$. This is a factor $\Th{P}$ less than the globally striped algorithm from Section~\ref{s:theoretical} can sort since \emph{every} PE must be able to hold one buffer block from each run in the merging phase.
However, $P\frac{m^2}{B}$ is $P$ times the amount that can be sorted by a single PE, which sounds very reasonable. 
In particular, any single PE equipped with a \emph{reasonable} amount of RAM and disks can sort the complete content of these disks in two passes since for technological reasons the price ratio between one byte of disk space and one byte of RAM has always been bounded by a few hundred. In this sense, the \CanonicalMergesort\ is sufficiently scalable.

The second most important restriction is that even the randomized algorithm
cannot move all the data to the right PE already during run formation. 
In Appendix~\ref{a:lb}, we show that this amount of data remains small
if $m\gg PB\log P$ (and the factor $\log P$ may be an artifact of the analysis), i.e., each PE must be able to store some number of blocks for each other PE.
This assumption is reasonable for the medium-sized machine we have used, and
for average case inputs, the $B$ disappears from the restriction, leading to an algorithm that scales even to very large machines with many thousands of PEs.
For very large machines and worst case inputs, our algorithm degrades to a three-pass algorithm which is still a good result.

A similar restriction on the local memory size applies to the external all-to-all algorithm from Section~\ref{ss:allall} -- each local memory must be able to hold a constant number of blocks for each other PE. However, randomization will mitigate this problem so that this part of the algorithm will scale to very large machines.

Another similar restriction applies to the multiway selection algorithm
described in Section~\ref{ss:select} whose naive implementation is only 
efficient if $m\gg PB\log M$. Here, our more clever implementation with
sampling and caching basically eliminates the problem.

\subsection{Further Issues\label{s:details}}

\paragraph*{Hierarchical Parallelism.}

In our terms, a PE is defined with respect to communication.
Data has to be communicated if and only if it resides on different PEs.
In practice, a PE can have multiple processors/cores and multiple disks.
We exploit this \emph{local} parallelism also.
The blocks on a PE are striped over the local disks.
For complex operations like internal sorting and merging, shared-memory parallel algorithms are used.
Thus, we exploit hierarchical parallelism.

Taking each processor core as a PE would lead to a larger number $P$, negatively influencing some of the stated properties of the algorithm.

\paragraph*{(Nearly) In-Place Operation.}

In the following, we analyze the additional external memory needed per node during the course of the algorithm.

The run formation in phase~1 can be done in-place easily.
The data is written back to the blocks where it was read from.

In phase~2, the multiway selection does not need considerable extra space.
The subsequent external all-to-all operation has a certain overhead since
in each suboperation it receives $P'$ pieces of data, which may lead to partially filled blocks. Since there is not sufficient internal memory to buffer all this data, these partially filled blocks have to be written out to disk.
Also, the in-place global external all-to-all needs $P + 1$ more blocks,
leading to a total temporary overhead of $RP'$ blocks per PE.

For merging in phase~3, blocks that are read to internal buffers are deallocated from disk immediately, so there are always blocks available for writing the output.

\paragraph*{Overlapping.}

For run formation, we overlap internal computation and communication with I/O.
While run $i$ is globally sorted internally, we first write the (already sorted) run $i-1$ before fetching the data for run $i+1$.

As a special optimization for inputs that fit into internal memory, we also overlap for this single-run case:
Immediately after a block is read from disk, it is sorted, while the disk is busy with subsequent blocks.
When all blocks are read and sorted, the algorithm only has to merge the blocks instead of still sorting everything.

We could also use overlapping of internal computation and communication in the internal global sort, splitting up the internal sort into three phases: local internal sort, global distribution, local internal merge.
However, our current implementation does not yet support this.
It is questionable whether this would help the performance, since all three operations still share the memory bandwidth.

\section{Implementation\label{s:implementation}}

We have implemented \CanonicalMergesort\ in C++.
We used the STXXL, the standard template library for XXL data sets~\cite{DKS08}, for handling asynchronous block-wise access to the multiple disks highly efficiently.
To sort and to merge data internally we used the parallel mode of the STL implementation of GCC 4.3.1, which exploits multi-core parallelism, and is based on the Multi-Core Standard Template Library~\cite{SSP07}.
Communication between nodes is done using the message passing interface MPI~\cite{mpi94}, we used MVAPICH~1.1 here.
Unfortunately, in MPI, data volumes are specified using 32-bit signed integers.
This means that no data volume greater than 2\,GiB can be passed to MPI routines.
We have re-implemented \texttt{MPI\_Alltoallv} to break this barrier.

\section{Experimental Results\label{s:experiments}}

The testing machine was a 200-node Intel Xeon cluster running Linux kernel 2.6.22.
Each node consists of two Quad-Core Intel Xeon X5355 processors clocked at 2\,667\,MHz with 16\,GiB main memory and 2$\times$4\,MiB cache.
The nodes are connected by a 288-port InfiniBand 4xDDR switch, the resulting point-to-point peak bandwidth between two nodes is more than 1\,300\,MB/s.
However, this value decreases when most nodes are used because the fabric gets overloaded (we have measured bandwidths as low as 400\,MB/s).
On every compute node, the 4 disks were configured as RAID-0 (striping)\footnote{Parallel disks are also directly supported by the program and could lead to even better timings, but we could not configure the machine accordingly at the time of testing.}.
Each node contains 4 Seagate Barracuda 7200.10 hard drives with a capacity of 250\,GB each.
We have measured peak I/O rates between 60 and 71\,MiB/s, in average 67\,MiB/s, on an XFS file system.
If not stated otherwise, we used a block size of 8\,MiB.
One cluster node corresponds to one PE in the following.

\begin{figure*}
\begin{minipage}[t]{0.48\textwidth}
\begin{center}
\includegraphics[width=\textwidth]{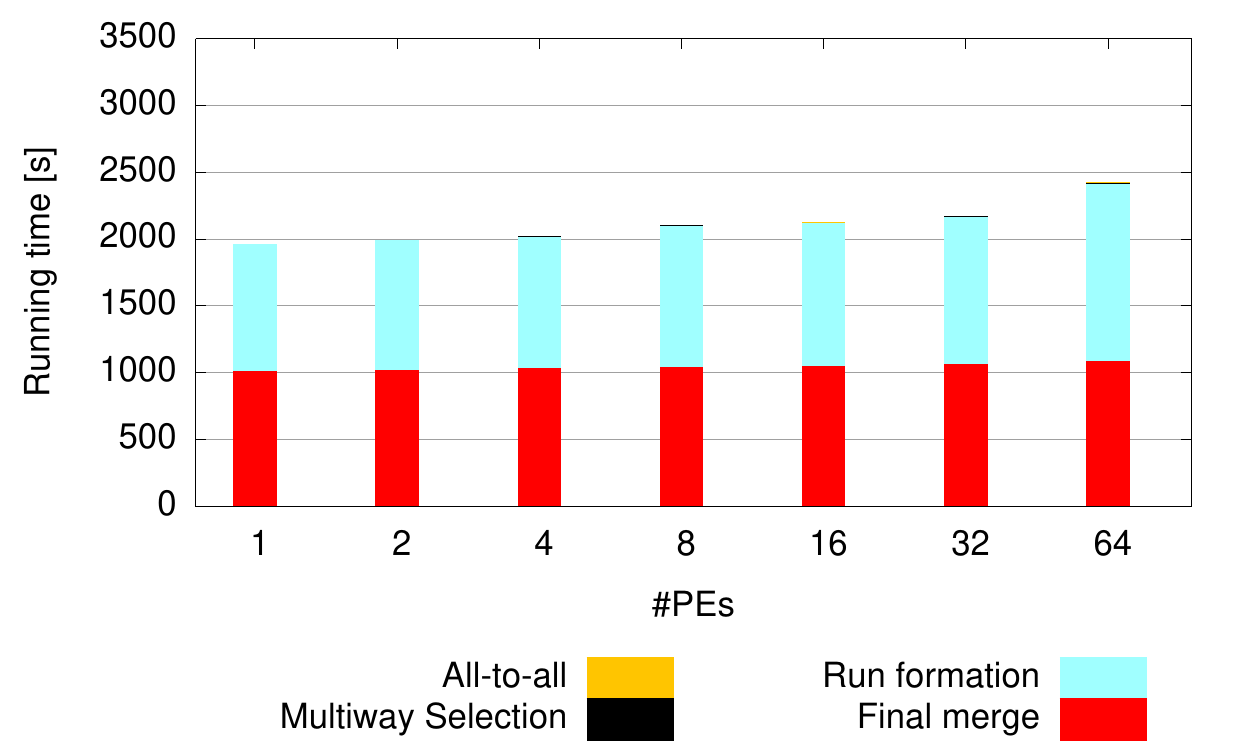}
\caption{\small \sf \textbf{Running times for random input, split up by the phases of the algorithm.
Please note that the order of the phases is different from the order in the algorithm, to allow for better visual comparison.}\label{f:rt_random}}
\end{center}
\end{minipage}
\hfill
\begin{minipage}[t]{0.48\textwidth}
\begin{center}
\includegraphics[width=\textwidth]{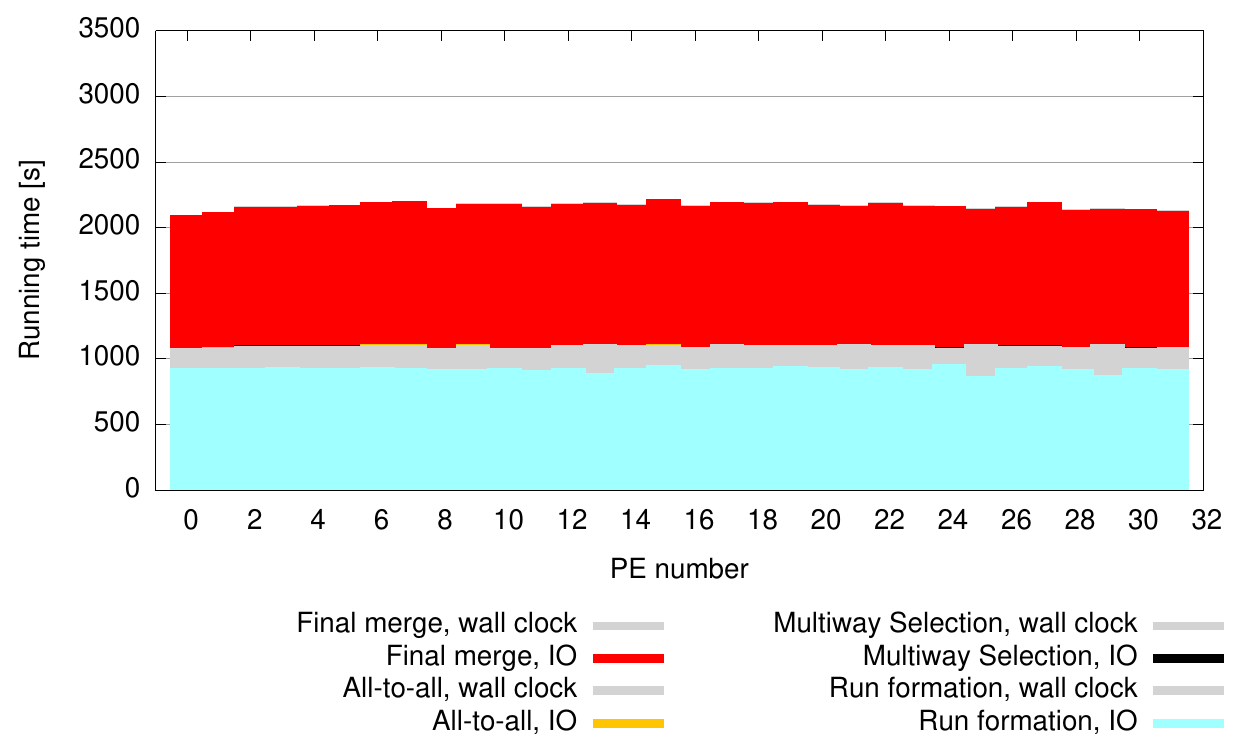}
\caption{\small \sf \textbf{Running times of the different phases on 32~nodes for random input.
For each phase, both the I/O time and the wall clock time are shown.
If there is a grey gap, the phase is not fully I/O-bound, as is the case for the run formation.}
\label{f:rt_phases_32}}
\end{center}
\end{minipage}

\vskip0.3cm

\begin{minipage}[t]{0.48\textwidth}
\begin{center}
\includegraphics[width=\textwidth]{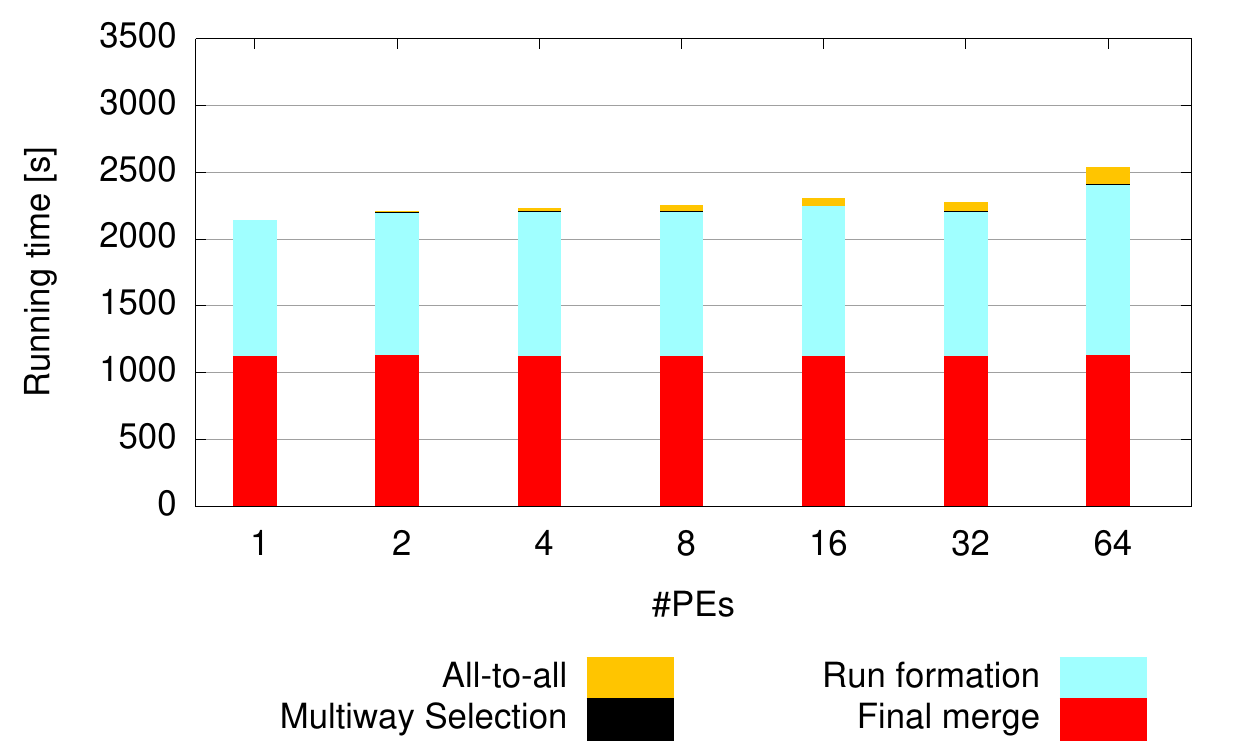}
\caption{\small \sf \textbf{Running times for worst-case input \emph{with} randomization applied.}\label{f:rt_wc_randomization}}
\end{center}
\end{minipage}
\hfill
\begin{minipage}[t]{0.48\textwidth}
\begin{center}
\includegraphics[width=\textwidth]{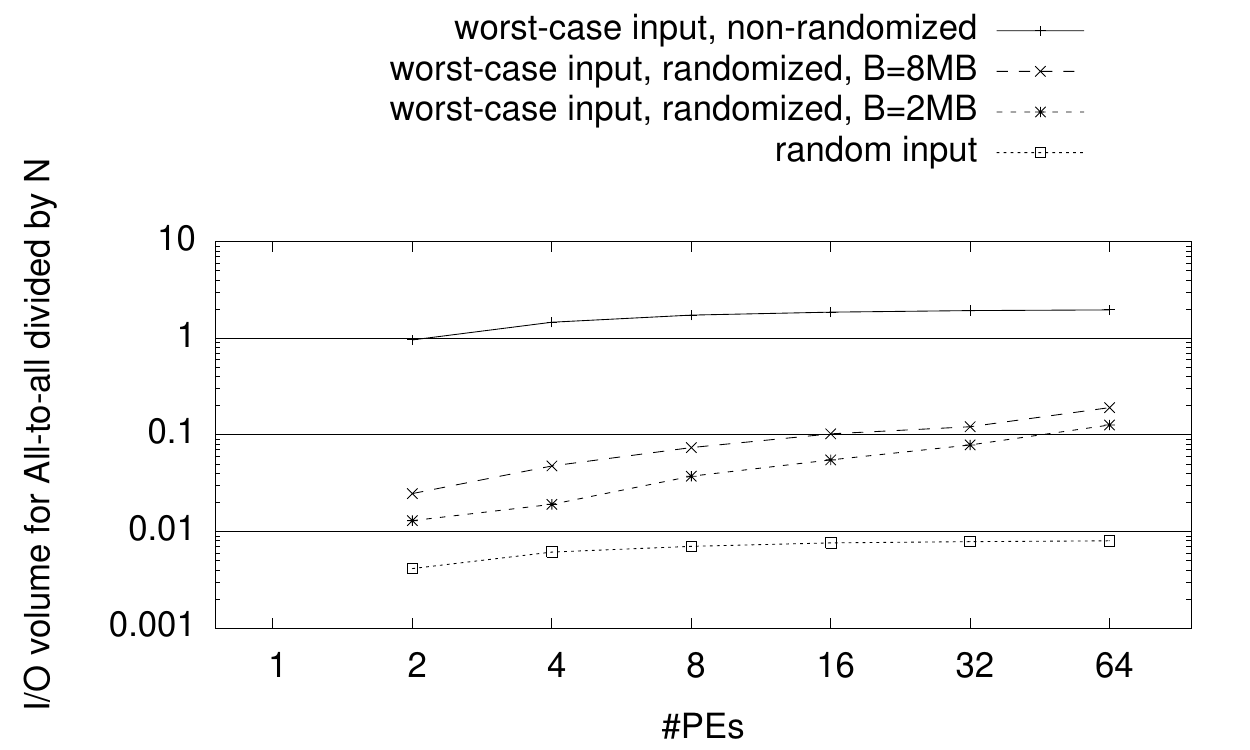}
\caption{\small \sf \textbf{I/O volume for the all-to-all phase for different inputs with/without randomization.}\label{f:iovolumes}}
\end{center}
\end{minipage}

\vskip0.3cm

\begin{minipage}[t]{0.48\textwidth}
\begin{center}
\includegraphics[width=\textwidth]{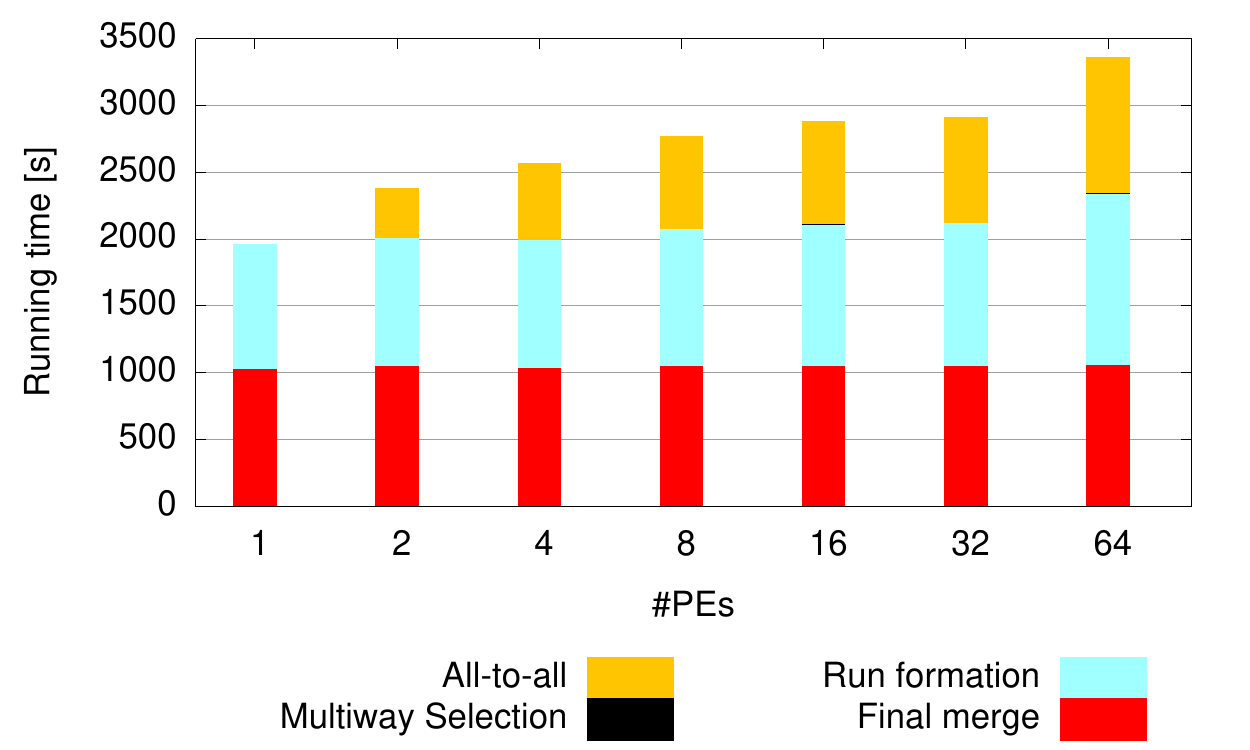}
\caption{\small \sf \textbf{Running times for worst-case input \emph{without} randomization applied.}\label{f:rt_wc_norandomization}}
\end{center}
\end{minipage}
\hfill
\end{figure*}

We tested scalability by sorting 100\,GiB of data per PE, with an increasing number of PEs, up to 64.
The element size is (only) 16~bytes with 64-bit keys. This makes internal computation efficiency as important as high I/O throughput.
As shown in Figure~\ref{f:rt_random}, the scalability is very good for random input data.
For worst-case input, a penalty of up to 50\% in running time can appear (Figure~\ref{f:rt_wc_norandomization}), as expected by the additional I/O performed by the all-to-all phase.
This overhead can be diminished by using randomization (Figure~\ref{f:rt_wc_randomization}), which reduces the I/O volume greatly.
Figure~\ref{f:iovolumes} shows that a smaller block size of 2\,MiB can further improve the effect of randomization, but usually at the cost of a little worse I/O performance.

As expected, run formation takes about the same time as the final merging.
The average I/O bandwidth per disk is about 50\,MiB/s, which is more than 2/3 of the maximum. The reasons for this overhead are worse performance of tracks closer to the center of a disk (when disks fill up), file system overhead, natural spreading of disk performance, and startup/finalization overhead.
Multiway selection takes in fact only negligible time.

Figure~\ref{f:rt_phases_32} shows the time consumption across the nodes for a 32-node run.
The work is very well balanced, but there is some variance in disk speed.

We have not compared our program to implementations of other algorithms directly. However, we made experiments on the well-established SortBenchmark, initiated by Jim Gray in 1984, and continuously adapted to the changing circumstances \cite{SortBenchmarkHomepage}.
This setting considers 100-byte elements with a 10-byte key.
The results~\cite{DEMSort} using 195 nodes show that we can sort $10^{12}$ bytes in less than 64~seconds, which is about a third of the time needed by the 2007 winner TokuSampleSort.
This is despite the fact the we use the same number of cores\footnote{For such large elements, the algorithm is not compute-bound at all.}, but only a third of the hard disks.
We also slightly improve on a recent result for the Terabyte category published informally by Google~\cite{GooglePetaSortBlog}, where 12\,000 disks\footnote{Google used 3-fold redundancy, but still, at least the performance of 4\,000 disks could be achieved.  Also, for a machine like ours, redundancy is not that desperately needed.} were used instead of 780 as in our case.

In the MinuteSort category, a time limit of one minute is given, the processed amount of the data is the metric.
We have beaten the former record of TokuSampleSort by a factor of 3.6, processing 955\,GB of data, which rendered the TerabyteSort category obsolete.
Yahoo achieved a result half as high using the Hadoop framework~\cite{YahooSortbenchmark2009,Hadoop}, but with a machine 7 times as large.

However, for the results mentioned so far, $N<M$, so the sort is merely internal and only 2 I/Os per block of elements are needed.

Concerning the newly established GraySort category, we can sort $10^{14}$ bytes (close to 100\,TiB) in slightly less than three hours on 195~nodes, resulting in about 564\,GB Bytes of sorted data per minute.
The Google program in this case takes only twice the time for ten times the amount of data, but they use an even larger machine than before, featuring 48\,000 disks, which is a factor of 61 larger.
The better performance of a factor of 5 is thus reduced to less than 0.1 in terms of relative efficiency.
Yahoo's official result of 578\,GB/min is only 2.5\% faster than us, but its efficiency is even worse, since they used 17 times the number of nodes.
Those nodes were very similar to the ones used by us, except having only half the memory.
They also had a worse communication bandwidth.
However, this would not have been a limiting factor for our algorithm.

\section{Conclusion\label{s:conclusion}}

We explored some of the design space for merging based parallel external sorting. Our globally striped algorithm minimizes the required I/Os.
Our \CanonicalMergesort\ algorithm is theoretically a bit less scalable but
it has close to minimal communication overhead, and a more useful output format.
Moreover, it sorts any technologically reasonable inputs in two to three passes.
For medium-sized machines or average case inputs, the I/O requirement remains closer to two passes than three passes.

A number of interesting questions remain for the future, in particular, a stronger analysis. Run formation could perhaps be improved to allow longer runs \cite[Section~5.4.1]{Knu98}. The main effect is that by decreasing the number of runs, we can further increase the block size. For the very largest inputs this could yield a slight improvement in performance. An interesting question is whether on large machines that have considerably higher communication bandwidth than I/O bandwidth, the globally striped algorithm could indeed be the better choice. This algorithm could also be useful for pipelined sorting where
the run formation does not \emph{fetch} the data but obtains it from some data generator (no randomization possible for \CanonicalMergesort) and where the output is not \emph{written} to disk but fed into a postprocessor that requires its input in sorted order (e.\,g., variants of Kruskal's algorithm \cite{OSS09}).
When scaling to very large machines, fault tolerance will play a bigger role.
An interesting question is whether this can be achieved with lower overhead than in \cite{GooglePetaSortBlog}.

\paragraph*{Acknowledgments.} Tim Kieritz provided an implementation of an early algorithm variant. Roman Dementiev's work on STXXL is an important reason for the efficiency of the code.

\newpage

%\bibliographystyle{plain}
%\bibliography{diss}

\input{dist_mem_ext_sort.bbl}
\newpage

\begin{appendix}
\subsection{Prefetching with Global Striping\label{a:theodetails}}

Perhaps the easiest way to do prefetching is to simply use the order in which
the blocks are needed for merging, which is determined by the order of the
smallest\footnote{In \cite{Knu98} and a lot of subsequent work, the
  \emph{largest} key in the \emph{previous block} is used. Here we use the
  approach from \cite{DemSan03} which is arguably more elegant and allows the
  merger to proceed for slightly longer without having to wait for a block from
  disk.}  keys in the data blocks. Unfortunately, it is open whether this leads
to optimal I/O rates unless $\Om{D\log D}$ prefetch buffer blocks are available
\cite{BarGroVit97}. However, in \cite{DemSan03}, very good performance is
observed for random inputs. In \cite{DemSan03,HutSanVit05}, an optimal
prefetching algorithm is used that is efficient already for $\Om{D}$ buffer
blocks.  This algorithm is based on
simulating a buffered writing process which iteratively fills a shared write
buffer and then simulates the output of one block on each disk whose queue contains
a write request.  We can use the sequential algorithm from \cite{HutSanVit05} as
long as $B=\Om{P}$ since we only perform constant work for each data block.
Parallelizing this algorithm is possible but requires relatively fine-grained
coordination: Allocating $\Oh{D}$ blocks to disk queues can be implemented
using plain message passing since randomization ensures that the number of
blocks per queue is at most $\Oh{\log D/\log\log D}$.  Simulating the outputs
can be done locally on each PE. But then we have to count how many write queues
are nonempty, which requires a global sum-reduction.  This is possible in time
$\Oh{\log P}$ on distributed-memory parallel machines.

\subsection{Scalable  Multiway Selection\label{app:select}}

Recall that during run formation, we sample every $K$-th element from each run together with its run number and position. For this analysis, we choose $K=B$, sample the smallest element of each block, and store it in a sample sequence together with its original position. Note that this approach is similar to the prediction sequences already mentioned in Section~\ref{s:theoretical}. To initialize multiway selection of the element of rank global $r$, we sort (in parallel) the sample, and extract the element $x$ of sample-local rank $\floor{r/B}$. By scanning the sorted sample backwards, we also find the predecessor of $x$ in each run. These elements are the initial values for the approximate splitter positions and $B$ is the initial step size.
This initialization can be done for all $P$ desired ranks using a parallel sorting step and a single parallel scan of the sorted sample and thus takes only a factor $\Oh{1/B}$ of the total work performed for sorting the full input.
Subsequently, the remaining selection work for PE $i$ only depends on one block in each run, i.\,e., overall, we have to access $RP$ blocks and deliver them to the PEs that have to process them. Randomization ensures that the number of blocks to be accessed and communicated on each PE is well balanced, i.\,e., close to the expectation $R$.
The final phase of multiway selection can work locally and in parallel and needs time $\Oh{R\log R \log B}$.

\subsection{Analysis of Data Redistribution\label{a:lb}}

We use the bounded difference inequality
\cite{McD89} which bounds how any function $f$ of independent random variables
is concentrated  around its mean, provided that changing a single variable
does not change the function value by too much. In our case we have $M/B$ random variables that are used to determine the blocks sorted by a particular run $j$.%
\footnote{Note that these block indices are \emph{not independent}. However, if we determine the blocks to be used by generating local random ppermutations the standard algorithm for determining random permutations uses $M/B$ independent random values for determining the blocks used in a single run.} The function $f$ we consider is the global rank of the smallest element of run $j$ that is stored on PE $i$. The expectation of $f$ is $iN/P$.
The deviation from this rank is proportional to the amount of data from run $j$ to be moved to PE $i$. By changing a single random variable,
the value of $f$ changes by at most the block size $B$. We get
\begin{equation}\prob{f-\expect[f]\geq t} \leq \exp\left(-\frac{t^2}{2\frac{N}{B}B^2}\right)=\exp\left(-\frac{t^2}{2MB}\right),\label{eq:bdineq}\end{equation}
i.\,e., it is unlikely that the more than $\Oh{\sqrt{MB}}$ elements have to be moved per run. However, we have to be a bit careful since the running time of a parallel algorithm depends on the PE where things are worst. Equation~(\ref{eq:bdineq}) also shows that it is unlikely that \emph{any} PE has to move more than $\Oh{\sqrt{MB\log P}}$ elements for run $j$. Since we are \emph{really} interested in the worst sum of data movements over all runs, the
truth lies somewhere in the middle. Anyway, $\Oh{R\sqrt{MB\log P}}$ is an upper bound for the expected amount of data movement to/from any PE. 
This is small compared to the total per PE amount of data movement of $N/P$
if $M/P\gg PB\log P$, i.\,e., each PE must be able to store $\Om{P\log P}$ data blocks (and the $\log P$ factor is probably overly conservative). This is a reasonable assumption for small and medium $P$ but does not hold for very large machines.%
\footnote{For example, our nodes can hold only about 2\,000 of the very large blocks we are using.} We also see that the reorganization overhead grows with the square-root of $B$ (Figure~\ref{f:iovolumes} supports this claim). Hence, on large machines, it might pay to use a smaller block size for reading blocks during run formation. Note that this affects only one fourth of the I/Os and we will furthermore not see the worst-case behavior of fully random accesses here since
during run formation, we can use offline disk scheduling techniques to reduce seek times and rotational delays.

\paragraph*{Average Case.} By setting $B=1$, we get a bound on the data movement for random inputs, i.\,e., for a random permutation of distinct elements. For low data movement, we need the condition that $M/P\gg P\log P$, i.\,e., every PE must be able to hold a logarithmic amount of data for every other PE in its local memory. This is a very mild condition. Indeed, internal memory parallel sorting algorithms that work with a single communication of the data have a similar condition.

\end{appendix}

\end{document}

%% file: makros.tex
\newcommand{\ceil}[1]{\left\lceil #1\right\rceil}
\newcommand{\floor}[1]{\left\lfloor #1\right\rfloor}

\newcommand{\realrange}[2]{\left[#1, #2\right]}

\newcommand{\unitrange}[2]{\realrange{0}{1}}

\newcommand{\prob}[1]{{\mathbb{P}}\left[#1\right]}

\newcommand{\expect}{{\mathbb{E}}}

\newcommand{\Oh}[1]{\mathcal{O}\!\left( #1\right)}

\newcommand{\oh}[1]{\mathrm{o}\!\left( #1\right)}
\newcommand{\Th}[1]{\Theta\!\left( #1\right)}
\newcommand{\Om}[1]{\Omega\left(#1\right)}

\newcommand{\llabel}[1]{\label{\labelprefix:#1}}
\newcommand{\labelprefix}{} % later redefined using renewcommand

\newcommand{\discussionsize}{\small}

\newcommand{\frage}[1]{{\sf #1}\marginpar{{\bf ?}}}

\newenvironment{code}{\noindent%\sf%
\begin{tabbing}%
\hspace{2em}\=\hspace{2em}\=\hspace{2em}\=\hspace{2em}\=\hspace{2em}\=%
\hspace{2em}\=\hspace{2em}\=\hspace{2em}\=\hspace{2em}\=\hspace{2em}\=%
\kill}{\end{tabbing}}

\newcommand{\labelcommand}{}
\newcommand{\captiontext}{}
\newsavebox{\codeparam}
\newcounter{lineNumber}
\newenvironment{disscodepos}[3]{%
\renewcommand{\labelcommand}{#2}%
\renewcommand{\captiontext}{#3}%
\sbox{\codeparam}{\parbox{\textwidth}{#3}}%
\begin{figure}[#1]\begin{center}\begin{code}\setcounter{lineNumber}{1}}{%
\end{code}\end{center}\caption{\llabel{\labelcommand}\captiontext}\end{figure}}

{\end{disscodepos}}

\newdimen\endofsize\endofsize=0.5em
\def\endofbeweis{~\quad\hglue\hsize minus\hsize
                 \hbox{\vrule height \endofsize width
\endofsize}\par}